\documentclass[reprint,showpacs,preprintnumbers,amsmath,amssymb,prc,floatfix]{revtex4-1}
\usepackage{color}
\usepackage{graphicx}% Include figure files
\usepackage{dcolumn}% Align table columns on decimal point
\usepackage{bm}% bold math
\usepackage[colorlinks,citecolor=blue,linkcolor=red,anchorcolor=blue,filecolor=blue,urlcolor=blue]{hyperref}

\begin{document}

%\preprint{AP/123-QED}
\preprint{XIV International Workshop on Hadron Physics}

\title{Exploring the decay probability of neutron-rich superheavy nuclei}

\author{M. Bhuyan}
\email{Email: bhuyan@ita.br}
\author{B. V. Carlson}
\email{Email: brettvc@gmail.com}
\affiliation{
Instituto Tecnol\'ogico de Aeron\'autica, 12.228-900 S\~ao Jos\'e dos Campos, S\~ao Paulo,
Brazil}
%\homepage{http://www.ita.br/}

\date{\today}

\begin{abstract}
The modes of decay for the even-even isotopes of superheavy nuclei of Z = 118 and 120 with neutron 
number $160 \leq N \leq 204$ are investigated in the framework of the axially deformed relativistic 
mean field model. The asymmetry parameter $\eta$ and the relative neutron-proton asymmetry of the 
surface to the center ($R_{\eta}$) are estimated for the ground state density distributions of the 
nuclei. We suggest that the resulting asymmetry parameter $\eta$ and the relative neutron-proton 
asymmetry $R_{\eta}$ of the density play a crucial role in the preformation factor of the decay 
half life. 
\end{abstract}

\pacs{PACS: 21.10.Dr, 21.60.Jz,23.60.+e, 27.90.+b}
\maketitle

Over the last three decades, the synthesis of superheavy nuclei has been dramatically rejuvenated 
owing to the emergence of cold fusion reactions, performed mainly at GSI, Darmstadt 
\cite{hof00,hof95,hof96,hof98,hof99}, and hot fusion and/or the actinide based fusion reactions 
performed mainly at JINR, Dubna \cite{oga98,oga01,oga04,oga07,oga10,oga11}. Through these advancements 
of stable nuclear beam technology, it is not only possible to synthesizes superheavy nuclei but also 
provide impressive prospects for understanding the nuclear properties of these nuclei 
\cite{hof00,hof95,hof96,hof98,hof99,oga98,oga01,oga04,oga07,oga10,oga11,eic07}. At present, the 
question of the mode of decay and the stability of these newly synthesized nuclei arises. While 
reviewing the production and decay properties of nuclei with atomic number $100 \leq Z \leq 118$, 
it can be seen that the sustainability of these superheavy nuclei is controlled mainly by the 
spontaneous fission and $\alpha$-decay processes 
\cite{hof00,hof95,hof96,hof98,oga98,oga01,oga04,oga07,oga10,oga11,eic07}. An important factor in 
the decay process of superheavy nuclei is the shell effect. It supplies the extra binding energy 
and increases the barrier height of fission \cite{sat04,pat91,poen06,sama07,delin07,zhang07}. The 
situation in the case of spontaneous fission is very complex as compared to the $\alpha$-decay 
process along the stability line of the superheavy region. Further, there is also the possible 
$\beta^-$-decay mode for the superheavy nuclei that proceeds via the weak interaction. This 
process is slower and less favored as compared to spontaneous fission and $\alpha$-decay in the 
valley of stability.

The most stable superheavy nuclei are predicted to be located along the neutron rich region of the 
$\beta$-stability line. It is not possible to reach those directly by the above mentioned fusion 
reactions with stable ion beams. In fact, the predicted magic proton number for the superheavy 
region are quite different within various theoretical approaches. For example, the magic proton 
number Z = 114 was predicted in the earliest macro-microscopic calculations \cite{meld66,nilson69}, 
and later confirmed by Refs. \cite{pat91,moll92}. Fully microscopic approaches predict the proton 
shell closure at Z = 120 \cite{bein74,sakeb12,bhu12}, and/or Z =126 \cite{cwiok96} using selected 
nucleon-nucleon interactions in mean field models. The neutron magic number N = 184 is almost 
firmly predicted by different theoretical models \cite{moll92,bhu12}. For further experimental 
study of the superheavy nuclei, especially near the neutron rich side of the nuclear chart, basic 
ideas of the internal structure and reaction mechanism of those nuclei from advanced theoretical 
approaches are required. In other words, in order to produce superheavy nuclei in the laboratory, 
one needs to know the internal configuration and the radioactive decay properties theoretically. 
Hence, the knowledge of the modes of decay and half-lives of a nucleus over a very wide range of 
neutron-proton asymmetry within advanced theories are essential for their synthesis process and 
further progress in experiments. 

In this regard, we investigate different possible modes of radioactive decay for the neutron rich 
superheavy nuclei. We have used the relativistic mean field (RMF) formalism \cite{bogu77,sero86} 
with the recently developed NL3$^*$ interaction parameter s\cite{plb09} for the present analysis. 
The model have been successfully applied in the description of nuclear structure phenomena both 
in $\beta-$stable and $\beta-$unstable regions throughout the nuclear landscape including superheavy 
nuclei \cite{bogu77,sero86,plb09,lala99,bhu09,bhu11,ring96,vret05,meng06,niks11,zhao12,bhu15,bhu15a}. 
The constant strength scheme is adopted to take into account pairing correlations \cite{estal01,bhu09} 
and evaluate the pairing gaps for neutrons and protons using the BCS equations \cite{pres75}. The 
aim of the present study is to determine the properties of the modes of decay of neutron rich 
superheavy nuclei, which may help us to answer some important open questions: (1) how far may we 
still move in synthesis of superheavy elements by the fusion reactions, (2) where the island of 
stability is centered, (3) what are the properties of the most stable superheavy nuclei, and (4) how 
can one reach to this region. Further, the decay properties also play a crucial role in the study of 
the $r$-process of nucleosynthesis as well as the formation of heavy and superheavy nuclei in nature 
\cite{pano09,gori15}. Here we have considered the isotopic chains Z = 118 and 120 with $160 \leq N 
\leq 204$, predicted to be the next magic valley \cite{patra00,sakeb12,bhu12} after $^{208}$Pb. The 
basic idea is that the decay process is highly influenced by the internal configuration of the 
nucleus. In other words, the internal arrangement of nucleons determines the stability and modes 
of decay of the nucleus. 
%%%%%%%%%%%%%%%%%%%%%%%%%%%%%%%%%%%%%%%%%%%%%%%%%%%%%%%%%%%%%%%%%%%%%%%%%%%%%%%%%%%%%%%%%%%%%%%%%%
\begin{table}
\caption{The RMF (NL3$^*$) result for binding energy, root-mean-square charge radius $r_{ch}$, the
quadrupole deformation parameter $\beta_2$, the energy difference between the ground and first
intrinsic excited state ($\Delta E$) and the relative neutron-proton asymmetry at the surface
to the center $R_{\eta}$ for $^{278-322}$118.}
\renewcommand{\tabcolsep}{0.2cm}
\renewcommand{\arraystretch}{1.0}
\begin{tabular}{cccccccccccc}
\hline
Nucleus & BE & $\beta_{2}$ & $r_{ch}$ & $\Delta$E & $R_{\eta}$ \\
\hline
$^{278}$118 & 1966.5 & 0.258&6.271 &0.258 & 2.51 \\
$^{280}$118 & 1983.4 & 0.553&6.495 &0.553 & 2.54 \\
$^{282}$118 & 2000.1 & 0.553&6.507 &0.553 & 2.58 \\
$^{284}$118 & 2016.1 & 0.554&6.520 &0.554 & 2.61 \\
$^{286}$118 & 2031.7 & 0.551&6.529 &0.551 & 2.64 \\
$^{288}$118 & 2046.8 & 0.543&6.535 &0.543 & 2.69 \\
$^{290}$118 & 2061.7 & 0.533&6.538 &0.533 & 2.71 \\
$^{292}$118 & 2076.3 & 0.528&6.546 &0.528 & 3.04 \\
$^{294}$118 & 2090.2 & 0.535&6.563 &0.535 & 3.16 \\
$^{296}$118 & 2103.5 & 0.544&6.582 &0.544 & 3.20 \\
$^{298}$118 & 2116.2 & 0.554&6.602 &0.554 & 3.33 \\
$^{300}$118 & 2128.3 & 0.564&6.624 &0.564 & 3.47 \\
$^{302}$118 & 2140.0 & 0.580&6.651 &0.580 & 3.61 \\
$^{304}$118 & 2151.2 & 0.582&6.667 &0.582 & 3.96 \\
$^{306}$118 & 2161.6 & 0.590&6.688 &0.590 & 4.11 \\
$^{308}$118 & 2171.6 & 0.609&6.721 &0.609 & 4.26 \\
$^{310}$118 & 2181.3 & 0.622&6.747 &0.622 & 4.41 \\
$^{312}$118 & 2192.8 & 0.753&6.893 &0.753 & 4.57 \\
$^{314}$118 & 2202.6 & 0.766&6.923 &0.766 & 4.71 \\
$^{316}$118 & 2212.1 & 0.776&6.950 &0.776 & 4.87 \\
$^{318}$118 & 2216.9 & 0.571&6.749 &0.571 & 4.94 \\
$^{320}$118 & 2225.6 & 0.525&6.720 &0.525 & 5.06 \\
$^{322}$118 & 2233.7 & 0.534&6.739 &0.534 & 5.13 \\
\hline
\end{tabular}
\label{table1}
\end{table}
%%%%%%%%%%%%%%%%%%%%%%%%%%%%%%%%%%%%%%%%%%%%%%%%%%%%%%%%%%%%
\begin{table}
\caption{The RMF (NL3$^*$) result for binding energy, root-mean-square charge radius $r_{ch}$, the   
quadrupole deformation parameter $\beta_2$, the energy difference between the ground and first
intrinsic excited state ($\Delta E$) and the relative neutron-proton asymmetry at the surface
to the center $R_{\eta}$ for $^{280-324}$120.}
\renewcommand{\tabcolsep}{0.20cm}
\renewcommand{\arraystretch}{1.0}
\begin{tabular}{cccccccccccc}
\hline
Nucleus & BE & $\beta_{2}$ & $r_{ch}$ & $\Delta$E & $R_{\eta}$ \\
\hline
$^{280}$120 & 1962.54 & 0.258 & 6.300 & 0.058 & 2.51 \\
$^{282}$120 & 1980.67 & 0.248 & 6.307 & 0.201 & 2.54 \\
$^{284}$120 & 1997.28 & 0.233 & 6.309 & 0.926 & 2.57 \\
$^{286}$120 & 2013.78 & 0.567 & 6.306 & 0.340 & 2.63 \\
$^{288}$120 & 2029.97 & 0.562 & 6.309 & 0.929 & 2.68 \\
$^{290}$120 & 2045.56 & 0.556 & 6.313 & 0.301 & 2.71 \\
$^{292}$120 & 2060.87 & 0.547 & 6.285 & 0.729 & 2.75 \\
$^{294}$120 & 2075.85 & 0.541 & 6.385 & 0.916 & 2.81 \\
$^{296}$120 & 2090.29 & 0.545 & 6.400 & 2.394 & 2.87 \\
$^{298}$120 & 2104.30 & 0.554 & 6.305 & 0.058 & 2.91 \\
$^{300}$120 & 2117.63 & 0.564 & 6.311 & 3.291 & 3.99 \\
$^{302}$120 & 2130.28 & 0.586 & 6.318 & 3.691 & 3.11 \\
$^{304}$120 & 2142.57 & 0.591 & 6.326 & 4.462 & 3.21 \\
$^{306}$120 & 2154.10 & 0.596 & 6.340 & 4.744 & 3.38 \\
$^{308}$120 & 2164.84 & 0.600 & 6.747 & 4.439 & 3.43 \\
$^{310}$120 & 2175.19 & 0.614 & 6.831 & 4.727 & 3.61 \\
$^{312}$120 & 2186.32 & 0.726 & 6.858 & 5.396 & 3.79 \\
$^{314}$120 & 2196.88 & 0.726 & 6.880 & 5.837 & 3.94 \\
$^{316}$120 & 2206.90 & 0.729 & 6.642 & 5.797 & 4.07 \\
$^{318}$120 & 2216.86 & 0.742 & 6.643 & 5.743 & 4.16 \\
$^{320}$120 & 2221.24 & -0.436& 6.634 & 0.707 & 4.27 \\
$^{322}$120 & 2230.56 & -0.445& 6.618 & 0.765 & 4.41 \\
$^{324}$120 & 2239.09 & -0.448& 6.606 & 0.544 & 4.63 \\
\hline
\end{tabular}
\label{table2}
\end{table}
%%%%%%%%%%%%%%%%%%%%%%%%%
\begin{figure}[ht]
%\vspace{0.65cm}
\begin{center}
\includegraphics[width=0.9\columnwidth,height=0.9\columnwidth]{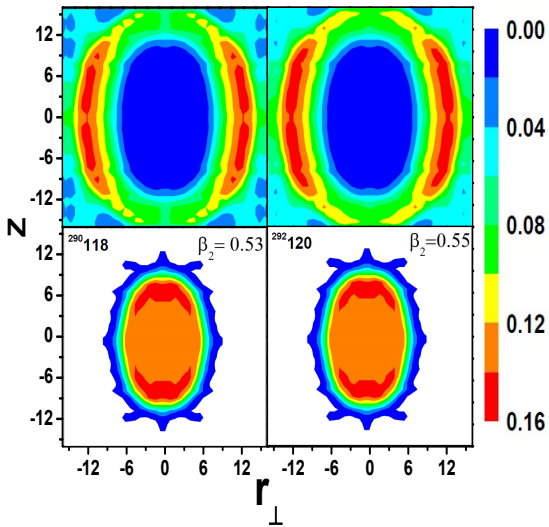}
\caption{\label{fig:1} (Color online) The total ground state density and the neutron-proton 
asymmetry parameter $\eta$ distribution for the density of $^{290}$118 and $^{292}$120 from 
the relativistic mean field (RMF) with NL3$^*$ force are displayed in the lower and upper 
panels, respectively.}
\vspace{-0.65cm}
\end{center}
\end{figure}
%%%%%%%%%%%%%%%%%%%%%%%%%%%%%%%%%%%%%%%%%%%%%%%%%%%%%%%%%%%%%%%%%%%%%%%%%%%%%%%%%%%%%%%%%%%%%%%
%%%%%%%%%%%%%%%%%%%%%%%%%%%%%%%%%%%%%%%%%%%%%%%%%%%%%%%%%%%%%%%%%%%%%%%%%%%%%%%%%%%%%%%%%%%%%%%

\begin{figure}[ht]
%\vspace{-0.65cm}
\begin{center}
\includegraphics[width=0.9\columnwidth,height=0.9\columnwidth]{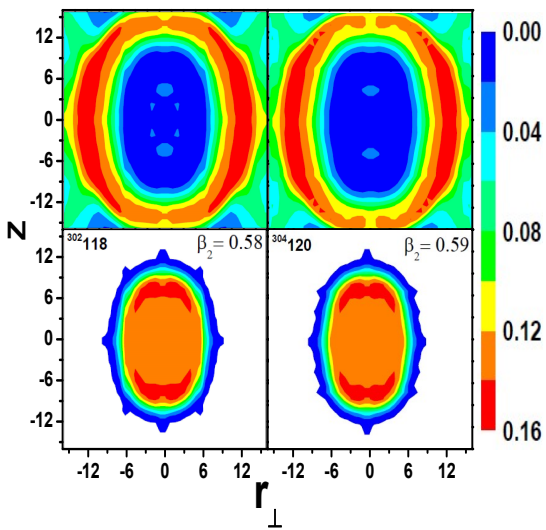}
\caption{\label{fig:2} (Color online) The total ground state density and the neutron-proton            
asymmetry parameter $\eta$ distribution for the density of $^{302}$118 and $^{304}$120 from 
the relativistic mean field (RMF) with NL3$^*$ force are displayed in the lower and upper 
panels, respectively.}
\vspace{-0.65cm}
\end{center}
\end{figure}

\begin{figure}[ht]
%\vspace{0.65cm}
\begin{center}
\includegraphics[width=0.9\columnwidth,height=0.9\columnwidth]{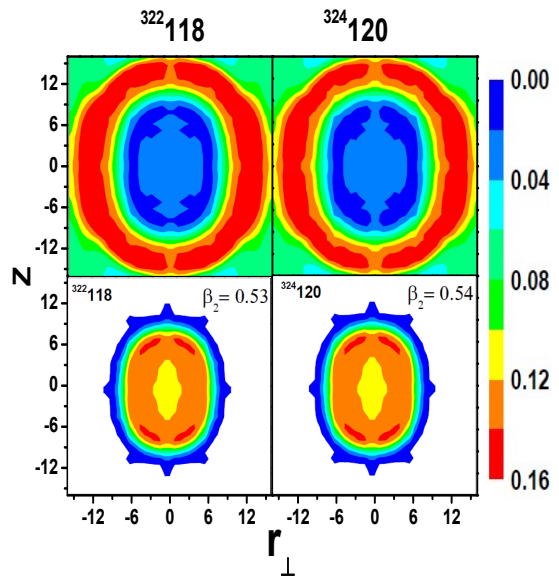}
\caption{\label{fig:3}(Color online) The total ground state density and the neutron-proton            
asymmetry parameter $\eta$ distribution for the density of $^{322}$118 and $^{324}$120 from 
the relativistic mean field (RMF) with NL3$^*$ force are displayed in the lower and upper 
panels, respectively.}
\vspace{-0.65cm}
\end{center}
\end{figure}

We have thus tried to explain the modes of decay of superheavy nuclei, by means of their internal 
structure and sub-structure. To know the proper internal configuration, it is important to know 
the ground and first intrinsic excited state properties of the nucleus. The bulk properties such 
as binding energy (BE), root mean square charge radii $r_{ch}$, matter radii, and the energy 
difference between the ground state and the intrinsic first excited state $\delta_E$ are calculated 
using the NL3$^*$ force parameter. The results for the isotopic chains of Z = 118 and Z = 120 are 
listed in Table. \ref{table1} and \ref{table2}, respectively. The quantity ($R_{\eta}$) in the last 
column of both the tables will be discussed in subsequent sections. The ground state solutions for 
the isotopic chains of Z = 118 and 120 have a deformed prolate configuration while the first excited 
states are found to be of spherical shape. An analysis of the internal structure of the nucleus is 
possible from the three dimensional (3D) contour plot of the deformed density. Here, we show the 
total density distribution, which is the sum of the proton $\rho_p$ and neutron $\rho_n$ density 
of the nucleus for the ground state solution. The total density of the nucleus extends from the 
center to a distance of about 7 $fm$, as seen in Figs. \ref{fig:1}, \ref{fig:2} and \ref{fig:3}.   

Due to their symmetry, the densities need only be obtained for the positive quadrant of the plane 
parallel to the {\it z-axis} (the symmetry axis), and are evaluated in the $r_{\bot}z-$plane, where 
$x$ = $r_{\bot} Cos\varphi$ and $y$ = $r_{\bot} Sin\varphi$ (cylindrical coordinates). The contour 
plots of the total density for the ground state of Z=118 and 120 for N = 172, 182 and 204 are shown 
in the lower panel of Figs. \ref{fig:1}, \ref{fig:2} and \ref{fig:3}, respectively. The color code 
along with the density ranges are given on the right side of the contour plots. From the color code, 
we can determine the range of the density values for a specific region of the nucleus (i.e. the 
cluster structures). For example, the color code with {\it deep red} corresponds to maximum density 
value ($\rho \sim$ 0.18 $fm^{-3}$) and the {\it deep blue} to the minimum density ($\rho \sim$ 0.001 
$fm^{-3}$). A careful inspection of the contour plots of the ground state density distributions shows 
a broken {\it ring}-like structure at the surface of the isotopes along the valley of stability. 
The distorted {\it ring}-like structure in the case of proton rich isotopes for N = 172 ($^{290}$118 
and $^{292}$120) (Fig. \ref{fig:1}) disappears as one moves towards the neutron rich isotopes (Fig. 
\ref{fig:2}). The asymmetry of the density in the {\it ring}-like region is higher than that of the 
central region of a nucleus. It shows a clear signature of a neutron skin structure for the neutron 
rich isotopes of Z = 118 and 120. This special attribute in the surface region of the nucleus also 
could play crucial role in the mode of decay of these nuclei.  
%%%%%%%%%%%%%%%%%%%%%%%%%%%%%%%%%%%%%%%%%%%%%%%%%%%%%%%%%%%%%%%%%%%%%%%%%%%%%%%%%%%%%%%%%%%
%%%%%%%%%%%%%%%%%%%%%%%%%%%%%%%%%%%
%%%%%%%%%%%%%%%%%%%%%%%%%%%%%%%%%%%%%%%%%%%%%%%%%%%%%%%%%%%%%%%%%%%%%%%%%%%%%%%%%%%%%%%%%%%%%%%

The asymmetry parameter $\eta$ can be estimated from the mean field density distributions and is 
defined as,
\begin{equation}
\eta = \frac{(\rho_n - \rho_p)}{(\rho_n + \rho_p)}.
\label{eta}
\end{equation}
Here, $\rho_n$ and $\rho_p$ are the density distribution of the proton and neutron, respectively. 
The role of this parameter is essential for estimation of the predominant constituents (neutron 
or/and proton) for a specific region. In other words, the contour plots of the asymmetry parameter 
$\eta$ using Eq. (\ref{eta}) shows the neutron-proton asymmetry distributions of the nucleus. The 
contour plots of $\eta$ for selected isotopes such as $^{290,302,322}$118 and $^{292,304,324}$120 
are shown in the upper panel of the Figs. \ref{fig:1}, \ref{fig:2} and \ref{fig:3}. The values of 
$\eta$ along with color codes are given to the right of the figure. From the figure, we find the 
distribution of the asymmetry parameter is uniform in the center of the nuclei but increases to 
a very high value (i.e. almost four times the central value) and forms a {\it ring}-like structure 
at the surface. Quantitatively, the central region (i.e. $\sim 0-5 fm$), has a value of $\eta \sim$ 
0.2, which changes to $\sim$ 0.8 at the surface region (i.e. $\sim 5-8 fm$). Note that that the value 
of $\eta$ is divided by 5 to use the color code in the right side of the figures. Here, we also find 
a {\it ring}-like structure for all the isotopes (Fig. \ref{fig:3}), but the situation is just 
reverse the of the ground state density (see Fig. \ref{fig:2}). The {\it ring}-like structures is 
also shifted toward the surface with increase of the neutron number. The concentration of neutrons 
in the surface region suggests a strong possibility for $\beta^-$-decay. Calculations in triaxially 
deformed coordinate space may resolve more issues and will throw more light on this possibility.  

We have estimated the relative neutron-proton asymmetry at the surface to that at the center of 
the nucleus $R_{\eta} = \eta_s/\eta_c$ (i.e. the ratio of the average asymmetries of the surface 
$\eta_s$ to the center $\eta_c$) for the isotopic chains of Z = 118 and Z = 120. The estimated 
values of relative neutron-proton asymmetry parameter $R_{\eta}$ are listed in the last column of 
Tables \ref{table1} and \ref{table2} for the Z = 118 and 120 isotopes, respectively. We find that 
the magnitude of $R_{\eta}$ increases with neutron number in each isotopic chain. The values of 
the $R_{\eta}$ is $\sim 2.5$ for $^{290}$118 (i.e. N=172) and increase gradually with the neutron 
number, reaching a value of $5.0$ for $^{322}$118. We see a similar tendency for the isotopic 
chain of Z = 120 (i.e. see Table \ref{table2}). The excess neutrons clearly accumulate in the 
surface region instead of the center of the nucleus. This effect is also manifested in the 
progressive appearance of a neutron {\it ring}-like structure with high neutron density. Due to 
the extreme neutron richness at the surface, $\beta^-$-decay could become a predominant mode of 
decay of neutron rich superheavy nuclei instead of $\alpha-$ decay. It might also be an alternative 
to the fission process of highly neutron rich nuclei \cite{sat04,sat10}, where the process is 
inhibited due to extreme neutron richness in the neck region. 

In summary, we have analyzed the bulk properties such as the binding energy, charge radius $r_{ch}$, 
and the binding energy difference of ground and intrinsic first excited states $\Delta_E$ for the 
isotopic chains of Z = 118 and 120. The RMF model, which has gained the confidence of the nuclear 
community in the study of exotic nuclei including superheavy nuclei has been adopted for the present 
study. We found deformed prolate ground state structure for these nuclei and a spherical first 
excited state. The ground state density along with the asymmetry parameter $\eta$ are also calculated. 
The widely varying relative neutron-proton asymmetry at the surface to that of the center of a 
nucleus $R_{\eta}$ suggests a possible $\beta^-$-decay mode. To our knowledge, this is one of the 
first such interesting and different phenomenon to appear in the superheavy region.    

This work is supported in part by FAPESP Project Nos. (2014/26195-5 \& 2017/05660-0), INCT-FNA 
Project No. 464898/2014-52014/26195-5, and CNPq-Brasil. The authors are also thankful to  S. K. 
Patra, Institute of Physics, for discussions and also for support throughout the work.


\begin{thebibliography}{00}
\bibitem{hof00}
S. Hofmann and G. M\"unzenberg, Rev. Mod. Phys. {\bf 72}, 733 (2000).
\bibitem{hof95}
S. Hofmann {\it et al.}, Z. Phys. A {\bf 350}, 277 (1995).
\bibitem{hof96}
S. Hofmann {\it et al.}, Z. Phys. A {\bf 354}, 229 (1996).
\bibitem{hof98}
S. Hofmann {\it et al.}, Rep. Prog. Phys. {\bf 61}, 639 (1998);
\bibitem{hof99}
S. Hofmann {\it et al.}, Acta Phys. Pol. B {\bf 30}, 621 (1999).
\bibitem{oga98}
Yu. Ts. Oganessian, Phys. Rev. Lett. {\bf 83}, 3154 (1998).
\bibitem{oga01}
Yu. Ts. Oganessian {\it et al.}, Nucl. Phys. A {\bf 685}, 17c (2001).
\bibitem{oga04}
Yu. Ts. Oganessian {\it et al.}, Phys. Rev. C {\bf 69}, 021601(R) (2004).
\bibitem{oga07}
Yu. Oganessian, J. Phys. G: Nucl. Part. Phys. {\bf 34}, R165 (2007).
\bibitem{oga10}
Yu. Ts. Oganessian {\it et al.}, Phys. Rev. Lett. {\bf 104}, 142502 (2010).
\bibitem{oga11}
Yu. Ts. Oganessian {\it et al.}, Phys. Rev. C {\bf 83}, 954315 (2011).
\bibitem{eic07}
R. Eichler {\it et al.}, Nucl. Phys. A {\bf 787}, 373c (2007).
\bibitem{sat04}
L. Satpathy and S.K. Patra, J. Phys. G {\bf 30}, 771 (2004).
\bibitem{pat91}
Z. Patyk and A. Sobiczewski, Nucl. Phys. A {\bf 533}, 132 (1991).
\bibitem{poen06}
D. N. Poenaru, I. H. Plonski, W. Greiner, Phys. Rev. C {\bf 74},
014312 (2006).
\bibitem{sama07}
C. Samanta, D. N. Basu, P. R. Chowdhury, J. Phys. Soc. Japan {\bf 76},
124201 (2007).
\bibitem{delin07}
D. S. Delin, R. J. Liotta, R. Wyss, Phys. Rev. C {\bf 76}, 044301 (2007).
\bibitem{zhang07}
H. F. Zhang, G. Royer, Phys. Rev. C {\bf 76}, 047304 (2007).
\bibitem{meld66}
H. Meldner, {\it Ph.D. thesis}, University Frankfurt am Main (1966).
\bibitem{nilson69}
S. G. Nilsson, C. F. Tsang, A. Sobiczewski, Z. Szymanski, S. Wycech, C. Gustafson, 
I.-L. Lamm, P. M\~oller, and B. Nilsson, Nucl. Phys., A {\bf 131}, 1 (1969).
\bibitem{moll92}
P. M\~oller and J. R. Nix, Nucl. Phys., A {\bf 549}, 84 (1992).
\bibitem{bein74}
M. Beiner, H. Flocard, M. V\`en\`eroni, and P. Quentin, Phys. Scr., {\bf 10A}, 84 (1974).
\bibitem{sakeb12}
S. Ahmad, M. Bhuyan, and S. K. Patra, Int. J. Mod. Phys. E {\bf 21}, 1250092 (2012).
\bibitem{bhu12}
M. Bhuyan, and S. K. Patra, Mod. Phys. Lett. A {\bf 27}, 1250173 (2012).
\bibitem{cwiok96}
S. \`Cwiok, J. Dobaczewski, P.H. Heenen, P. Magierski, W. Nazarewicz, Nucl. Phys., A 
{\bf 611}, 211 (1996).
\bibitem{bogu77}
J. Boguta and A. R. Bodmer, Nucl. Phys. A {\bf 292}, 413 (1977).
\bibitem{sero86}
B. D. Serot and J. D. Walecka, in {\it Advances in Nuclear Physics}, edited by J. W.
Negele and Erich Vogt {\it Plenum Press, New York}, Vol. {\bf 16}, p. 1 (1986).
\bibitem{plb09}
G. A. Lalazissis, S. Karatzikos, R. Fossion, D. Pena Arteaga, A. V. Afanasjev, and P.
Ring, Phys. Lett. B {\bf 671}, 36 (2009).
\bibitem{lala99}
G. A. Lalazissis, S. Raman and P. Ring, Atm. Data. Nucl. Data. Table. {\bf 71}, 1 (1999).
\bibitem{bhu09}
S. K. Patra, M. Bhuyan, M. S. Mehta and Raj K. Gupta, Phys. Rev. C {\bf 80}, 034312 (2009).
\bibitem{bhu11}
M. Bhuyan, S. K. Patra, and Raj K. Gupta, Phys. Rev. C {\bf 84}, 014317 (2011).
\bibitem{ring96}
P. Ring, Prog. Part. Nucl. Phys. {\bf 37}, 193 (1996).
\bibitem{vret05}
D. Vretenar, A. V. Afanasjev, G. A. Lalazissis, and P. Ring, Phys. Rep. {\bf 409}, 101 (2005).
\bibitem{meng06}
J. Meng, H. Toki, S. G. Zhou, S. Q. Zhang, W. H. Long, and
L. S. Geng, Prog. Part. Nucl. Phys. {\bf 57}, 470 (2006).
\bibitem{niks11}
T. Niksi\"c, D. Vretenar, and P. Ring, Prog. Part. Nucl. Phys. {\bf 66}, 519 (2011).
\bibitem{zhao12}
Xian-Feng Zhao, and Huan-Yu Jia, Phys. Rev. C {\bf 85}, 065806 (2012).
\bibitem{bhu15}
M. Bhuyan, S. K. Patra, and Raj K. Gupta, J. Phys. G: Nucl. Part. Phys.,
{\bf 42}, 015105 (2015).
\bibitem{bhu15a}
M. Bhuyan. Phys. Rev. C {\bf 92}, 034323 (2015).
\bibitem{estal01}
M. Del Estal, M. Centelles,  X. Vi\~nas and S. K. Patra, Phys. Rev. C {\bf 63}, 044321(2001); 
\bibitem{patra01a}
M. Del Estal, M. Centelles, X. Vi\~nas and S.K. Patra, Phys. Rev C {\bf 63}, 024314 (2001).
\bibitem{pres75}
M. A. Preston and R. K. Bhaduri, {\it Structure of the nucleus}, (Addition-Wesley Publishing Co.), 
p-144 (1975).
\bibitem{pano09}
I. V. Panov, I. Yu. Korneev, and F.-K. Thielemann, Phys. Atm. Nucl., {\bf 72}, 1026 (2009).
\bibitem{gori15}
S. Goriely, G. M. Pinedo, Nucl. Phys. A {\bf 944}, 158 (2015). 
\bibitem{patra00}
S. K. Patra, W. Greiner and R. K. Gupta, J. Phys. G: Nucl. Part. Phys. {\bf 26}, L65 (2000).
\bibitem{sat10}
S. K. Patra, R. K. Choudhury, and L. Satpathy, J. Phys. G: Nucl. Part. Phys. 
{\bf 37}, 085103 (2010).

\end{thebibliography}
\end{document}